# Single-shot laser-induced switching of an exchange biased antiferromagnet

Zongxia Guo[1,3#], Junlin Wang[4#], Gregory Malinowski[3], Boyu Zhang[1*], Wei Zhang[1,2], Hangtian Wang[1,3], Chen Lyu[1,3], Yi Peng[3], Pierre Vallobra[1,2], Yongbing Xu[4,5], Sarah Jenkins[5], Roy W. Chantrell[5], Richard F. L. Evans[5], Stéphane Mangin[3], Weisheng Zhao[1,2*] and Michel Hehn[3*]

[1]Fert Beijing Institute, School of Integrated Science and Engineering, Beihang University, Beijing, China.

[2]Anhui High Reliability Chips Engineering Laboratory, Hefei Innovation Research Institute, Beihang University, Hefei, China

[3]Université de Lorraine, CNRS, IJL, Nancy, France

[4]School of Integrated Circuits, Guangdong University of Technology, China

[5]School of Physics, Engineering and Technology, University of York, York, United Kingdom

[#]These authors contributed equally to this work

*Email : boyu.zhang@buaa.edu.cn,

weisheng.zhao@buaa.edu.cn,

michel.hehn@univ-lorraine.fr

# Abstract

Ultrafast manipulation of magnetic order has challenged our understanding the fundamental and dynamic properties of magnetic materials. So far single shot magnetic switching has been limited to ferrimagnetic alloys and multilayers. In ferromagnetic (FM)/antiferromagnetic (AFM) bilayers, exchange bias ($H_e$) arises from the interfacial exchange coupling between the two layers and reflects the microscopic orientation of the antiferromagnet. Here we demonstrate the possibility of single shot switching of the antiferromagnet (change of the sign and amplitude of $H_e$) with a single femtosecond laser pulse in IrMn/CoGd bilayers. We demonstrate the switching for a wide range of fluences for different layer thicknesses and compositions. Atomistic simulations predict ultrafast switching and recovery of the AFM magnetization on a timescale of 2 ps. These results provide the fastest and the most energy-efficient method to set the exchange bias and pave the way to potential applications for ultrafast spintronic devices.

**Introduction**

Exchange bias is the shift of the hysteresis loop along the magnetic field direction, named the Exchange Bias field ($H_e$), which typically arises from a Ferromagnetic (FM) layer in direct contact with an Antiferromagnetic (AFM) layer in an exchange-coupled FM/AFM system[1]. Exchange bias is an interface effect, where the value of $H_e$ depends mostly on layer thickness and texture, the magnetic configuration in both FM and AFM layers[2] and the procedure used to set up the interface and volume AFM configuration. Initializing exchange bias is a complex process and is done by heating the system above the Néel temperature $T_N$ or at least the blocking temperature $T_B$ of the AFM layer and cooling down with the FM layer in a saturated state, with or without an applied magnetic field[3,4]. This thermal annealing can be done in a quasi-static way (annealing for several hours), using rapid annealing or continuous laser sweeping on the surface[5,6].

Recent studies have demonstrated that spin currents can be used to manipulate exchange bias[7,8,9,10] thanks to spin-orbit torque (SOT) without any applied magnetic field. All-optical switching has also been used to manipulate exchange-biased systems combining heat and magneto-optical fields[11,12], but so far no single-shot switching has been reported. By using all-optical helicity-dependent switching (AO-HDS), the repeated reversal of a [Co/Pt] multilayer leads to the switching of the perpendicular exchange bias in [Co/Pt]$_n$/IrMn[13] after multiple laser pulses where the effect is explained by thermal annealing of the interface during the FM magnetization reversal. Ferrimagnets such as GdFeCo and CoGd are known to show the all-optical helicity independent switching (AO-HIS) by a single femtosecond laser excitation[14,15], which could further compress the timescale of the exchange bias switching.

In this article, we demonstrate the control and switching of the exchange bias in IrMn/CoGd antiferromagnetic/ferrimagnetic (AFM/FiM) heterostructures with perpendicular magnetic anisotropy using one single femtosecond laser pulse. We find that single-shot switching occurs for a wide range of laser fluence, compositions and layer thicknesses, within a timescale of 100 ps. Numerical simulations suggest an ultrafast demagnetization of IrMn in only 100 fs followed by an ultrafast recovery driven by the mutual recovery of the Co and Gd sublattice magnetizations. The ultrafast IrMn dynamics leads to a large reduction in the switched exchange bias and a probabilistic switching process for each grain.

**Stacks structure and single-shot exchange bias switching**

Fig. 1a presents a sketch of the multilayer stacks studied: Glass/Ta(5)/Pt(5)/Ir$_{20}$Mn$_{80}$($t_{IrMn}$)/Co$_{77}$Gd$_{23}$(4)/Pt(5) (thicknesses in nm) with $t_{IrMn}$ = 2 to 10 nm. The Co$_{77}$Gd$_{23}$ alloy is Co-dominant and the compensation concentration of Gd at room temperature is between 30% and 32% (see Supplementary Section 1). The perpendicular magnetic anisotropy and exchange bias of the IrMn/CoGd bilayer are confirmed by the polar magneto-optical Kerr effect (MOKE) measurements (Fig. 1b). The exchange bias is set up by annealing at 200°C during 1h with a 60 mT field applied perpendicular to film plane. The extracted exchange bias field $\mu_0 H_e$ and coercivity $\mu_0 H_c$ as a function of the IrMn thickness are presented in Fig. 1c. The evolution of $\mu_0 H_e$ with $t_{IrMn}$ is similar to that reported in the literature[16]. A noticeable exchange bias field can be observed for an IrMn thickness of $t_{IrMn}$=2 nm, reaching a maximum at $t_{IrMn}$=4 nm with $J_{ex}$= 0.16 mJ/m$^2$ (close to values reported in the literature[2]) and decreases when $t_{IrMn}$ is further increased. The coercivity $\mu_0 H_c$ decreases with increasing $t_{IrMn}$.

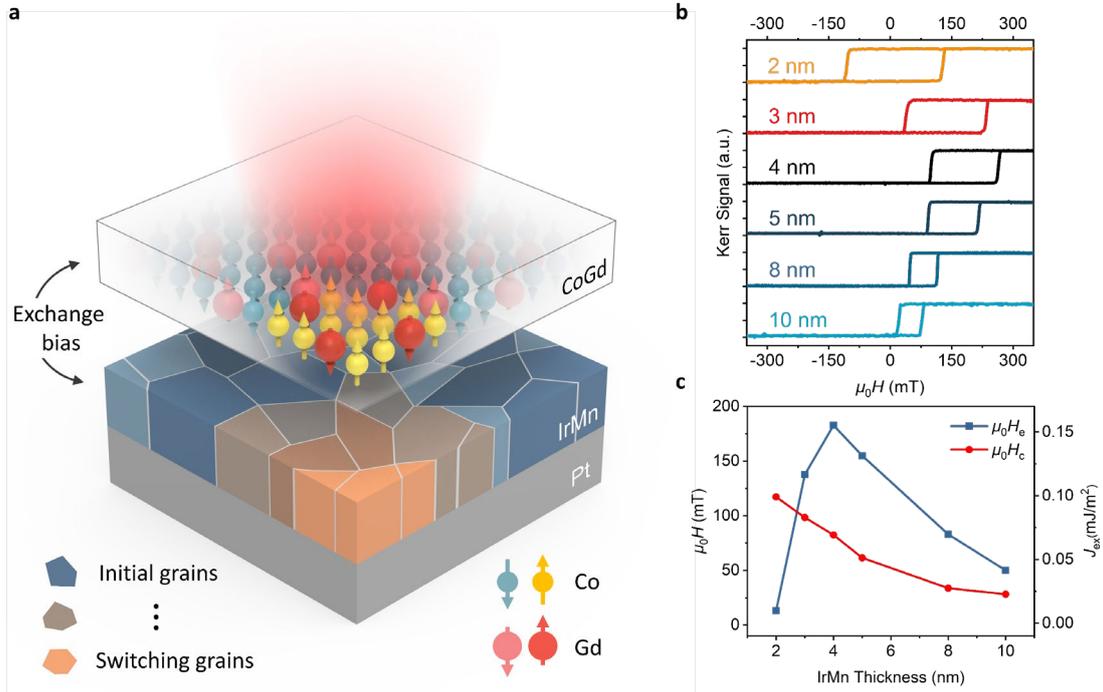

**Fig. 1 | IrMn/CoGd bilayer magnetic properties characterization. a,** Sketch of the film stack and the laser incident beam. **b,** Hysteresis loops obtained on the annealed stacks for different IrMn thickness ($t_{IrMn}$ from 2 to 10 nm) measured by polar MOKE. **c,** Exchange bias field $\mu_0 H_e$, effective exchange coupling constant $J_{ex}$ given by $J_{ex} = \mu_0 H_e M_s t_{CoGd}$, and coercivity $\mu_0 H_c$ as a function of the IrMn thickness ($t_{IrMn}$).

First, single-pulse laser switching experiments have been performed by exciting an IrMn(5)/Co$_{77}$Gd$_{23}$(4) sample from the top CoGd surface (Fig. 1a). Using a single laser pulse of 40 fs with a fluence of 17 mJ/cm$^2$, indues a reversal of the exchange field from 154 mT to -26 mT, as reported in Fig. 2a. Considering that the beam fluence follows a spatial Gaussian profile, a distribution of the exchange bias field along the radius of the laser spot is observed, as illustrated in Supplementary Section 3. The value of -26 mT was obtained at the center of the spot. In comparison to the initial exchange bias, the intensity of $\mu_0 H_e$ is strongly reduced, but interestingly, the sign is reversed. From this state, adding laser pulses leads to a change of $\mu_0 H_e$ sign at each pulse while the intensity remains almost constant as shown in Fig. 2b.

We then investigated the laser fluence and IrMn thicknesses dependences of the induced exchange bias modification. (Kerr images of CoGd switching can be found in Supplementary Section 5). As shown in Fig. 2c, independently of the IrMn thickness for $t_{IrMn}$ between 4 and 10 nm, a rapid reduction of $\mu_0 H_e$ is observed at low fluences (related $\mu_0 H_c$ is shown in Supplementary Section 4). It then reverses and increases for medium fluence excitation before saturating or even decreases towards zero at high fluence. First of all, the change of the $\mu_0 H_e$ sign can be closely related to the AO-HIS of CoGd magnetization by a single pulse (dashed line in Fig. 2c). Indeed, the switching of the CoGd is a precondition for reversing the interfacial magnetization that imprints the interfacial spin configuration in the IrMn during cooling. Second, starting from positive exchange bias obtained using thermal annealing under an applied magnetic field, the intensity of the maximum negative exchange bias observed after one pulse is always smaller. Depending on the IrMn thickness, 13% to 33% of the initial exchange bias field can be obtained with $t_{IrMn}$ decreasing from 10 to 4 nm respectively (see Supplementary Section 6).

The two-temperature model included in the atomistic simulations explains the ultrafast dynamics process for the IrMn/CoGd system. The simulation results show that, after the laser heats the sample, the temperature of the nanoscale IrMn grains exceeds the Néel temperature (around 700 K) and switches according to the AO-HIS of the CoGd layer. After the lattice temperature cools down to below the blocking temperature, the incomplete exchange bias reversal (Fig.2c) indicates that under such an ultrafast switching process, the recovery of the AFM order is not fully deterministic and thus does not allow the spin structure to rearrange fully. The thicker the IrMn layer is, the smaller the amplitude of the reversed exchange bias field is. This arises due to the weaker effective exchange field on the antiferromagnet that it is harder to orient the IrMn grains. To allow the spin structure to rearrange thermally, we induced heating using low fluence laser pulses, with fluences much lower than the threshold fluences needed to reverse the CoGd magnetization. After the CoGd switching with a single laser pulse of 17 mJ/cm$^2$ on the IrMn(5)/Co$_{77}$Gd$_{23}$(4) sample, low fluence laser pulses of 8.2 mJ/cm$^2$ were applied on the switched region. As the pulse number increases, regardless of whether CoGd magnetization and exchange bias field are in parallel (P) or antiparallel (AP) states, the exchange bias recovers to ±76 mT along the CoGd magnetization direction as illustrated in Fig. 2d (related coercivity $\mu_0 H_c$ is illustrated in Supplementary Section 7). The same behavior is observed for the 8 nm IrMn sample (see Supplementary Section 8). The prolonged heating generated by low fluence multi-pulses allows the AFM layer spins to further reorder and increase $\mu_0 H_e$. However, this process still does not allow reaching the initial exchange bias because the cooling time after the laser pulse is around 500 ps compared to the long annealing time and applied field, both of which allow the development of a larger exchange bias field through relaxation over energy barriers at equilibrium.

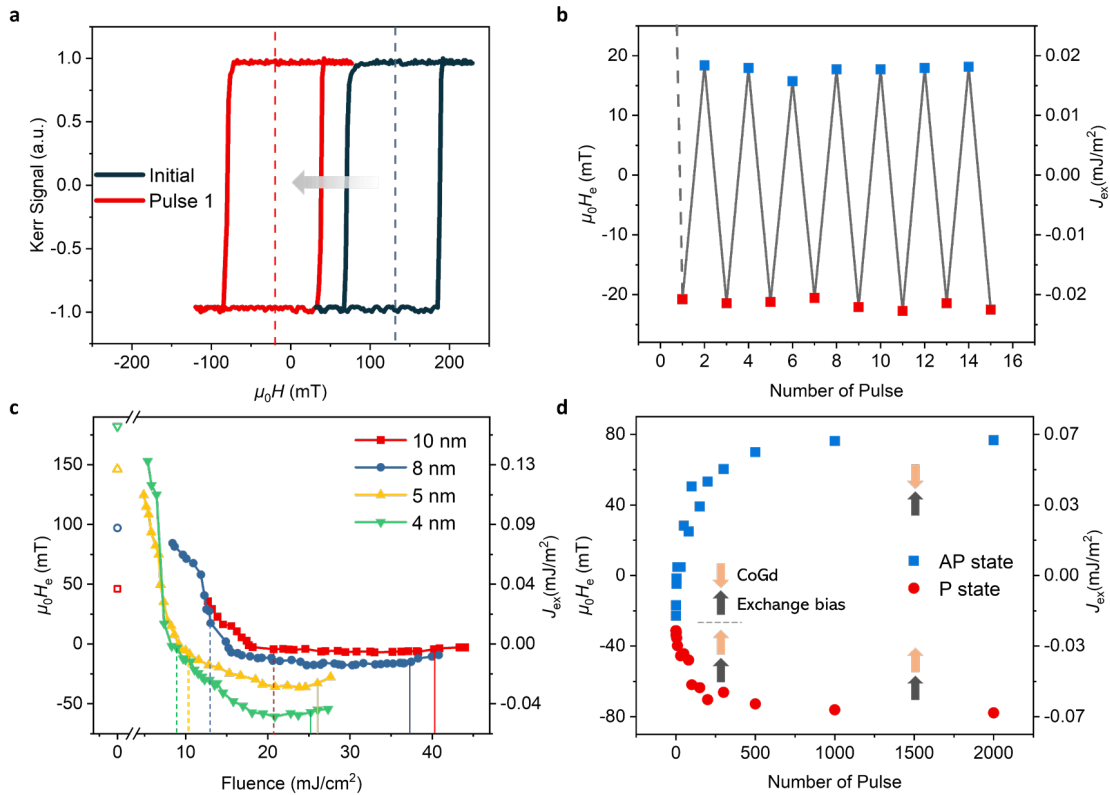

**Fig. 2 | Evolution of the exchange bias using a single femtosecond laser pulse. a,** Hysteresis loop of IrMn(5)/Co$_{77}$Gd$_{23}$(4) before and after exposure to a single linearly polarized laser pulse with a pulse duration of 40 fs and a fluence of 17 mJ/cm$^2$ **b,** Modulation of the exchange bias field as a function of the number of pulses with a pulse duration of 40 fs and a laser fluence of 17 mJ/cm$^2$. Starting from a bilayer showing an exchange bias field ($\mu_0 H_e$= 154 mT) and a coercivity ($\mu_0 H_c$= 61.5 mT) **c,** Evolution of $\mu_0 H_e$ after one single laser pulse as a function of the laser fluence. The dashed line indicates the CoGd switching threshold, and the solid line indicates the demagnetization threshold. The hollow symbols at 0 mJ/cm$^2$ indicate the initial $\mu_0 H_e$ of each IrMn thickness. **d,** Evolution of the $\mu_0 H_e$ as a function of the number of laser pulses. After the exposure of a single laser pulse, the switching region is illuminated by multi-pulse with a fluence of 8.2 mJ/cm$^2$ and a repetition rate of 5-kHz.

## Exchange bias switching

Since exchange bias can be manipulated using a single femtosecond laser pulse, the next question that needs to be answered is how long it takes to modify the exchange bias field and, consequently, the IrMn magnetic configuration at the interface? Several reports have suggested that photoexcitation of the AFM/FM interface induces significant modulations in the exchange bias field on ultrashort timescales[17,18,19]. Detailed time-resolved studies of dynamics showed that the characteristic timescale of laser-induced exchange bias quenching in a polycrystalline Co/IrMn bilayer is 0.7 ± 0.5 ps, attributing the rapid decrease in exchange coupling to a spin disorder at the interface created by laser heating[17].

In our case, the demagnetization and switching dynamics after a 40 fs laser pulse illumination of the annealed IrMn(5)/Co$_{77}$Gd$_{23}$(4) sample have been studied as a function of time for various laser fluences (Fig. 3a) and for different applied fields intensities (Fig. 3b). A Co demagnetization time of ~ 700 fs could be measured, independently of the applied field. We observe that for pump laser fluence larger than 11.7 mJ/cm$^2$, the thermally induced magnetization switching is observed in CoGd alloy[14,15]. The full reversal requires a typical time of ~ 5 ps. Switching dynamics show similar features in the other samples with different IrMn thicknesses (see Supplementary Section 9). Since a constant magnetic field is applied to reset the sample magnetization state in the TR-MOKE measurement, the presence of IrMn does not seem to change the dynamics of the ferrimagnetic layer as it was already reported in the case of NiMn layer[18]. At long timescales, the CoGd layer switches back to its initial magnetization orientation faster because of the presence of an external magnetic field increases. Fig. 3b shows that the larger the magnetic field, the faster the CoGd switches back, which can be explained using the time evolution of a macrospin state under applied field[20]. We can especially note that for an applied field of -220 mT, it takes ~500 ps for the magnetization to cross zero again towards positive magnetization after achieving a full reversal.

The single pulse evolution of the exchange bias has been studied as a function of the laser fluence for different applied fields as shown in Fig. 3c. We note that, due to the combination of laser heating and CoGd switching back under the negative applied field, the exchange bias field begins to recover at the larger laser fluences. By comparing the CoGd dynamics at the corresponding external field in Fig. 3b, we can deduce the longest timescale where the exchange bias can be set. For low fluence and up to 16 mJ/cm$^2$ (see point A in Fig. 3c), the exchange bias is independent of the applied field in the field range +150 mT/-240 mT. Since we can observe that at those fields, the reversed CoGd state is preserved reversed up to 100 ps, we can conclude that exchange bias is switched and set within those 100 ps. When increasing the fluence, the bilayers is more strongly affected by the applied field as can be seen in Fig. 3c. For point B (18 mJ/cm$^2$ with -150 mT), the exchange bias is increased due to thermal heating of the antiferromagnet above the blocking temperature and cooling in an applied field, thus resetting some of the switched grains. For fluences larger than 24 mJ/cm$^2$ (point C), the exchange bias field no longer increases under zero applied field and even tends towards zero as the fluence increases. At those fluences, the CoGd no longer switches but demagnetizes as suggested by the state diagram measured by Wei et al.[21]. Consequently, a demagnetized state is imprinted in the IrMn at the interface leading to exchange bias close to zero. However, with an applied field, either positive or negative exchange bias can be reached depending on the sign of the applied field, similarly to conventional exchange bias setting.

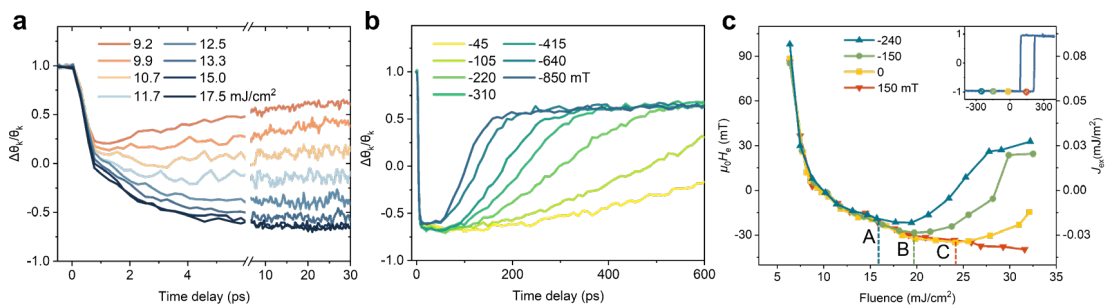

**Fig. 3 | The CoGd switching and exchange bias dynamic under an external field in IrMn(5)/Co$_{77}$Gd$_{23}$(4). a,** CoGd dynamics as a function of time for different laser fluences under an

applied field of -220 mT. The demagnetization time is ~ 700 fs and the magnetization switching time is ~ 5 ps. **b,** CoGd dynamics for a laser fluence of 17.5 mJ/cm$^2$ under various external fields. **c,** Evolution of $\mu_0 H_e$ after one single laser pulse as a function of the laser fluence under various external fields. The inset shows the hysteresis loop and marks the applied external field of IrMn(5)/Co$_{77}$Gd$_{23}$(4) sample.

**Dynamics of IrMn and exchange bias from atomistic simulations**

To better understand the laser-induced dynamics of IrMn and exchange bias switching, we perform atomistic simulations of the coupled IrMn/CoGd system including the granular structure of the film[22,23], as shown in Fig. 4a. The exchange bias is set using a field cooling procedure (see Supplementary Section 11). Fig. 4b shows the dynamic response of the Gd, Co and Mn sublattice moments to a 40 fs laser pulse simulated using the two-temperature model, showing ultrafast demagnetization at three distinct timescales. The difference in ferromagnetic demagnetization rates is due to the different magnetic moments and effective damping constants, but the demagnetization of the antiferromagnet is even faster than a ferromagnet with a characteristic timescale of only 15 fs, despite the larger moment of 2.6 $\mu_B$ / Mn site. The ultrafast dynamics in antiferromagnets is due to the high frequency optical spin-wave modes available due to the strong antiferromagnetic exchange coupling. In Fig. S12 and Fig S13 we show the role of ultrafast heating on the system, showing demagnetization and thermally induced switching of the CoGd and ultrafast disordering of the antiferromagnet, followed by a slow recovery of the total IrMn magnetization parallel to the Co magnetization, while each grain remagnetizes to a single domain state in only 2 ps. Due to the rapid heating the IrMn grains enter a superparamagnetic regime where the orientations of the four distinct antiferromagnetic sublattices fluctuate on the sub-picosecond timescale immediately after the laser pulse. Fig. 4c shows the time-dependent magnetization for the IrMn sublattice set in the perpendicular $\pm z$ direction by the field cooling procedure, in the perpendicular $\pm z$ direction for 5 characteristic grains (the data for all grains is shown in Supplementary Fig. S14). Each grain has a random coupling to the Co and Gd sublattices, meaning that around half of the grains are set parallel to the Co, and half are set parallel to the Gd magnetization. Switching is indicated by a full perpendicular re-orientation of the IrMn after the pulse. Due to the magnetic symmetry of $\gamma$-IrMn partial reorientations are also possible, with a z-component $S_z \sim \pm 0.86$, shown schematically in Fig. 4d. At short timescales, the IrMn demagnetizes in the first 0.2 ps, and then fluctuates strongly, while some grains fully switch after 0.5 ps following the Co/Gd magnetization respectively. The simulations indicate a complex process for IrMn switching that is probabilistic (Supplementary Table 2), the reduction of the exchange bias after the laser pulse. This is due to ultrafast thermal annealing, where the IrMn grains are heated above the Néel temperature then experience superparamagnetic relaxation above the blocking temperature of the grains. The resetting of the exchange bias by cooling through the blocking temperature in the exchange field from the Co is not significant on the sub-nanosecond timescales of the experiment as seen through the need of multiple pulses in Fig. 2(d). Therefore, the short timescale coherent switching process of the IrMn must play an important role in the single shot switching seen experimentally. However, the large heating after the pulse partially overwrites the switching process causing a less than 100% switching probability, but this is independent of the number of pulses and simply follows the magnetization of the Co/Gd sublattices. Essentially each pulse acts as a reset for the IrMn magnetization where the remagnetization follows the direction of the Co/Gd, but there are no progressive effects as the IrMn is completely demagnetized after each pulse.

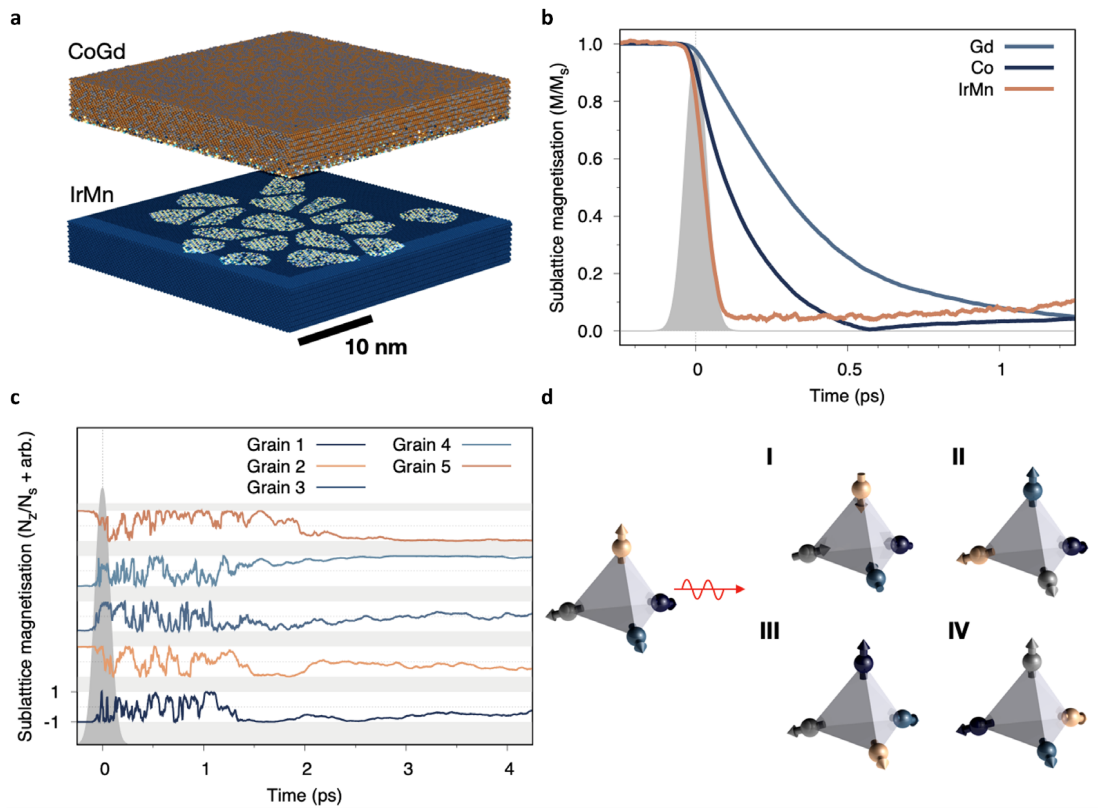

**Fig. 4 | Atomistic simulations of the dynamics of IrMn/CoGd bilayers. a,** Visualisation of the simulated system $30 \times 30 \times 10$ nm showing the granular nature of the IrMn. **b.** Ultrafast demagnetization

of the Gd, Co and IrMn sublattices, showing distinct demagnetization times for each sublattice. **c.** Individual response of the z-component of the IrMn sublattice magnetization of each grain showing a coherent switching process 1.5-2 ps after the pulse, before achieving a stable state at longer timescales. The sublattice chosen for each grain is the one that is set in the perpendicular direction ($\pm z$). **d.** Schematic of the switching process, showing reorientation of the Mn sublattice responsible for the exchange bias. In γ-IrMn3 there are four distinct sublattices, and switching of the antiferromagnet can occur by a full 180 degree reorientation (I) or by a 109.5 degree reorientation (II, III, IV) into one of the other three orientations with opposite z-orientation $S_z \sim 0.86$. This partial reorientation is one of the reasons for a reduced perpendicular exchange bias.

**Conclusion**

We have studied the effect of a single laser pulse on a model perpendicularly exchange bias system IrMn/CoGd bilayers. We have shown that the CoGd magnetization switches in the presence of an interfacial exchange coupling with the IrMn. The value and sign of the exchange bias can be controlled with a single femtosecond laser pulse showing single-shot reorientation of the antiferromagnet. Moreover, only using purely thermal heating, exchange bias can also be modified even with laser fluence below the CoGd switching threshold. Dynamic experiments indicate that the single femtosecond laser-induced exchange bias switching can be done within 100 ps. Atomistic simulations show the ultrafast demagnetization and coherent switching dynamics of the IrMn grains on the ultrafast timescale, switching in less than 2 ps. Our results demonstrate the first single-shot switching and control of an exchange biased antiferromagnet expanding the fundamental understanding and application aspects of exchange bias. The IrMn/CoGd system allows ultrafast, field-free and energy-efficient control of an exchange biased antiferromagnet with high ordering temperature and thermal stability, making it highly suited to applications.

**Methods**

**Sample preparation and magnetic properties of the films.** The film stacks mentioned in the paper were deposited on the glass substrate by DC magnetron sputtering. The CoGd alloy layers were deposited by co-sputtering Co and Gd targets. According to the static hysteresis loops measured with a polar magneto-optic Kerr looper (MOKE), we can note that the magnetization compensation point is between y = 30% and 32% at room temperature (300 K). All the deposited stacks were annealed at 200 °C for 1 hour in an out-of-plane saturation field of -60 mT to set the initial exchange bias. The interfacial exchange constant $J_{ex}$ was calculated using: $J_{ex} = \mu_0 H_e M_s t_{CoGd}$, $M_s$ is obtained by SQUID-VSM.

**Experimental set-up for all-optical switching and TR-MOKE.** We use a Ti: sapphire femtosecond laser with a 5-kHz repetition rate and a wavelength of 800 nm (1.55 eV). Both experiment platforms share the same laser source. In the all-optical switching set-up, the pulse duration is 40 fs and a half-wave (λ/2) plate role is to tune the laser fluence needed for the experiment. Then the lens focuses the beam on the sample such that the full width at half maximum (FWHM) of beam diameter is ~50 μm. The response of the magnetic film was studied using a polar MOKE microscope behind the sample to image the magnetic domains while the laser is shining on the sample. Magnetic hysteresis measurements of the illumination region were performed using a commercial Evico MOKE microscope with a white-color LED source. In the time-resolved measurement, the pump and probe are issued from the same laser. We use a beam splitter to split the laser into a strong (pump) and a weak (probe) pulse of 40 fs. The diameter of the pump laser is 470 μm and the probe laser is 70 μm. Time-resolved measurements are made by adjusting the delay between the pump and the probe lasers with a 0~1 ns range and a 2 fs minimum step size.

**Atomistic simulations.** We use an atomistic spin model to simulate the equilibrium and dynamic properties of the IrMn/CoGd bilayer. The energetics of the system are described by a spin Hamiltonian of the form

$$H = -\sum_{i<j} J_{ij} S_i \cdot S_j - \sum_i k_u S_i^z - \sum_i k_N (S_i \cdot e_{ij})^2 - \sum_i \mu_i S_i^z B^z$$

Where $S_{i,j}$ is a unit vector describing the direction of spin $i,j$, $J_{ij}$ is the exchange energy between atoms $i,j$, $k_u$ is the uniaxial anisotropy on the Co and Gd sites, $k_N$ is the Néel pair anisotropy between the Mn sites[24] and $B^z$ is the external applied magnetic field. The model parameters are listed in Supplementary Table 1. The equilibrium properties and field cooling are simulated with an adaptive Monte Carlo metropolis model with adaptive move[25] and dynamics are computed using the stochastic Landau-Lifshitz-Gilbert equation with Langevin dynamics[26]. The temperature rise induced by the laser pulse is modelled by coupling the spin system to the two-temperature model[27] augmented with a cooling term to reproduce the cooling effect of the substrate. The simulations were performed using the VAMPIRE software package[26].

**Acknowledgements**

This work is supported by the ANR-15-CE24-0009 UMAMI and the ANR-20-CE09-0013, by the Institute Carnot ICEEL for the project "Optic-switch" and Matelas, by the Région Grand Est, by the Metropole Grand Nancy, by the impact project LUE-N4S, part of the French PIA project "Lorraine Université d'Excellence," reference ANR-15-IDEX-04-LUE, and by the "FEDERFSE Lorraine et Massif Vosges 2014-2020," a European Union Program. The authors gratefully acknowledge the National Natural Science Foundation of China (Grant No. 12104031, 12104030, 61627813), the Program of Introducing Talents of Discipline to Universities (Grant No. B16001), the Beijing Municipal Science and Technology Project (Grant No. Z201100004220002), China Postdoctoral Science Foundation (Grants No.2022M710320) and China Scholarship Council. The atomistic simulations were performed on the Viking Cluster, a high-performance compute facility provided by the University of York and ARCHER2, the UK National Supercomputing Service (https://www.archer2.ac.uk).

**Author contributions**

S.M., WS.Z. and M.H. initialized, conceived and supervised the project. M.H. deposited the film stacks and M.H. and Y.P. optimized the magnetic and switching properties of CoGd. Z.G., G.M., W.Z. and T.H. performed the measurements. J.W., R.W.C and R.F.L.E performed the atomistic simulations. S.J prepared the input parameters for the IrMn/CoGd bilayer simulations. Z.G., B.Z., G.M., J. W., R.F.L.E, S.M. and M.H. wrote the manuscript. All authors discussed the results and commented on the manuscript.

**Competing interests**

The authors declare no competing interests.

# Supplementary materials for

# Single-shot laser-induced switching of an exchange biased antiferromagnet

Zongxia Guo[1,3#], Junlin Wang[4#], Gregory Malinowski[3], Boyu Zhang[1*], Wei Zhang[1,2], Hangtian Wang[1,3], Chen Lyu[1,3], Yi Peng[3], Pierre Vallobra[1,2], Yongbing Xu[4,5], Sarah Jenkins[5], Roy W. Chantrell[5], Richard F. L. Evans[5], Stéphane Mangin[3], Weisheng Zhao[1,2*] and Michel Hehn[3*]

[1]Fert Beijing Institute, School of Integrated Science and Engineering, Beihang University, Beijing, China.

[2]Anhui High Reliability Chips Engineering Laboratory, Hefei Innovation Research Institute, Beihang University, Hefei, China

[3]Université de Lorraine, CNRS, IJL, Nancy, France

[4]School of Integrated Circuits, Guangdong University of Technology, China

[5]School of Physics, Engineering and Technology, University of York, York, United Kingdom

[#]These authors contributed equally to this work

*Email : boyu.zhang@buaa.edu.cn,

weisheng.zhao@buaa.edu.cn,

michel.hehn@univ-lorraine.fr

**Section 1. Exchange bias field with various Gd concentrations.**

The film stack of Sub/Ta (5)/Pt (5)/Ir$_{20}$Mn$_{80}$ (4)/Co$_{100-y}$Gd$_y$ (4)/Pt (5) (nm) with y =23 to 32% were grown by DC magnetron sputtering on glass substrates. In the CoGd amorphous alloy, the magnetization of the Gd sublattice is antiferromagnetically exchange coupled to the magnetization of the Co sublattice. The net magnetization of the alloy is given by the contribution of both sublattices and reaches zero at the compensation point. We have characterized IrMn(4)/Co$_{100-y}$Gd$_y$(4) samples with various concentrations. The hysteresis loops are measured by polar MOKE (Fig. S1a). The summarized coercivity $\mu_0 H_c$ and exchange bias field $\mu_0 H_e$ are shown in Fig. S1b. According to the static hysteresis loops with a changed sign, we can note that the magnetization compensation point is between y = 30% and 32% at room temperature (300 K).

According to Meiklejohn's model[1], in a macrospin FM/AFM bilayer, the exchange bias field is evaluated using $\mu_0 H_e = J_{ex}/M_s t_{CoGd}$, where $J_{ex}$ is the effective exchange coupling constant, $M_s$ is the saturation magnetization of CoGd and $t_{CoGd}$ is the CoGd layer thickness. The good crystallization of IrMn (Supplementary Section 2) and the low $M_s$ of near compensated CoGd alloy lead to a strong exchange bias field compared to conventional IrMn/FM bilayers[1]. In addition, under the same field annealing conditions, the exchange bias field is opposite near the compensation point, which indicates that the interfacial AFM pinned spin is aligned with the magnetization of the Co moment[1].

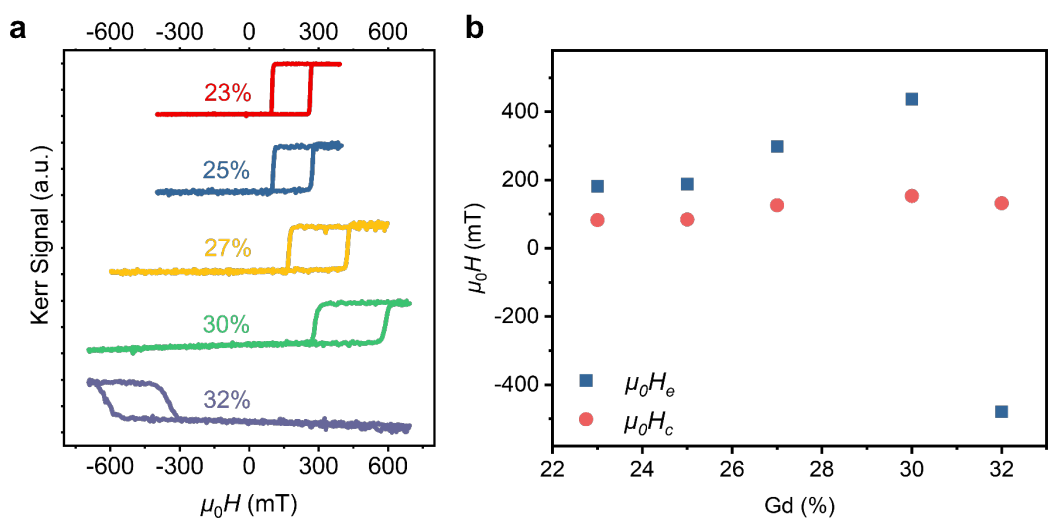

**Fig. S1 | a,** The hysteresis loops with various Gd concentrations (from 23% to 32%). **b,** Evolution of exchange bias field $\mu_0 H_e$ and coercivity $\mu_0 H_c$ as a function of Gd concentration.

**Section 2. X-ray diffraction patterns of Glass/Ta (5)/Pt (5)/Ir$_{20}$Mn$_{80}$ (x)/Co$_{77}$Gd$_{23}$ (4)/Pt (5).**

Magnetic properties of the IrMn based exchange bias system were generally correlated with the IrMn <111> texture. Therefore, the choice of seed layers for the IrMn layer becomes crucial[2]. The crystallographic structure of the samples after annealing was analyzed by the X-ray diffraction (XRD) technique shown in Fig. S2. Two XRD peaks can be obtained: the higher one belongs to the (111)-oriented Pt seed layer, and the other is (111)-oriented IrMn. The intensity of the IrMn (111) peak depends on the IrMn thickness.

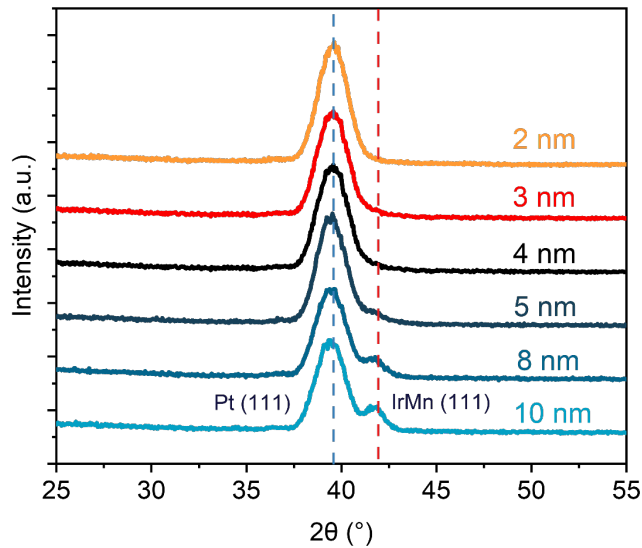

**Fig. S2** | The (111) texture peak of IrMn layer and Pt seed layer was carried out by θ-2θ XRD scans of the bilayer stacks for different IrMn thicknesses ($t_{IrMn}$ from 2 to 10 nm).

**Section 3. Exchange bias field and coercivity distribution within a laser spot.**

The switching of CoGd and exchange bias are closely related to the temperature gradient induced by Gaussian distributed laser fluence. Fig. S3 shows the evolution of the exchange bias field (Fig. S3a) and the coercive field (Fig. S3b) of IrMn (5)/Co$_{77}$Gd$_{23}$ (4) bilayer as a function of the distance from the spot center for various laser fluence. Before switching, the exchange bias rapidly reduces in the spot center; over the switching threshold of 11.25 mJ/cm$^2$, the exchange bias sign changes.

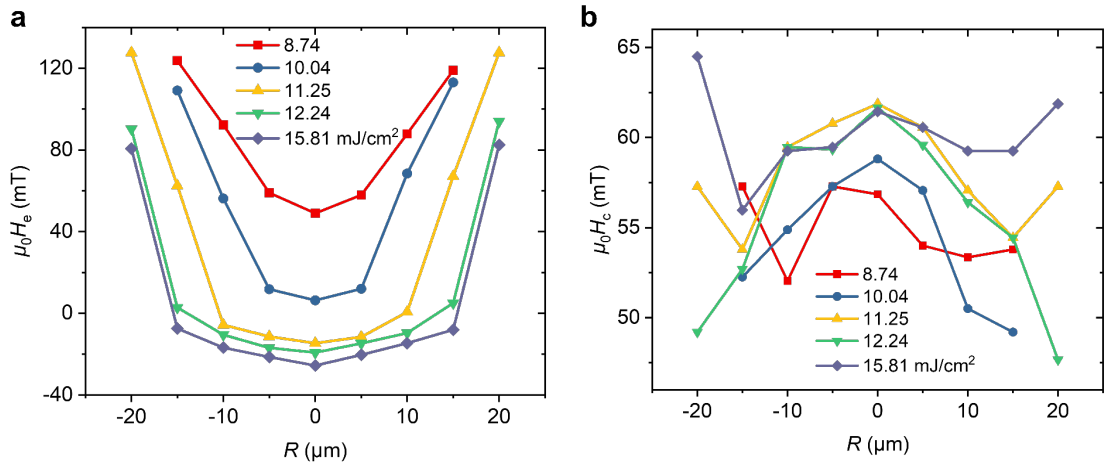

**Fig. S3 | a-b,** Exchange bias field $\mu_0 H_e$ and coercivity $\mu_0 H_c$ as a function of the position (R) from the centre of the laser spot at various laser fluences.

**Section 4. Evolution of the coercivity after a single laser pulse as a function of fluence.**

The coercivity change is the most often observed in the exchange bias system, which also reflects the properties of the interface coupling[3]. Fig. S4 shows the evolution of the coercivity $\mu_0 H_c$ of IrMn (5)/Co$_{77}$Gd$_{23}$ (4) bilayer after single laser pulse as a function of the laser fluence. We can note that in samples with thinner IrMn layers, the reduction in coercivity due to switching of the exchange bias is more pronounced.

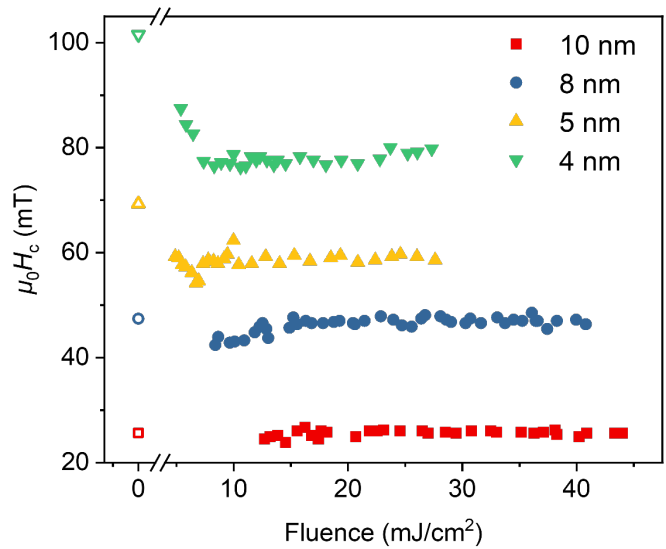

**Fig. S4 |** Evolution of the coercivity $\mu_0 H_c$ as a function of the laser fluence. The hollow symbols at 0 mJ/cm$^2$ indicate the initial $\mu_0 H_c$ of each IrMn thickness.

**Section 5. Kerr images of AO-HIS with various laser fluence.**

Fig. S5 shows the AO-HIS Kerr images of IrMn (5)/Co$_{77}$Gd$_{23}$ (4) bilayer with various laser fluences. The light grey part is the unswitched region with an initial exchange bias. The black circle is the switched region shined by a single femtosecond laser pulse. We can note that the switching threshold is 11.9 mJ/cm$^2$ and become a multidomain state over 29.4 mJ/cm$^2$.

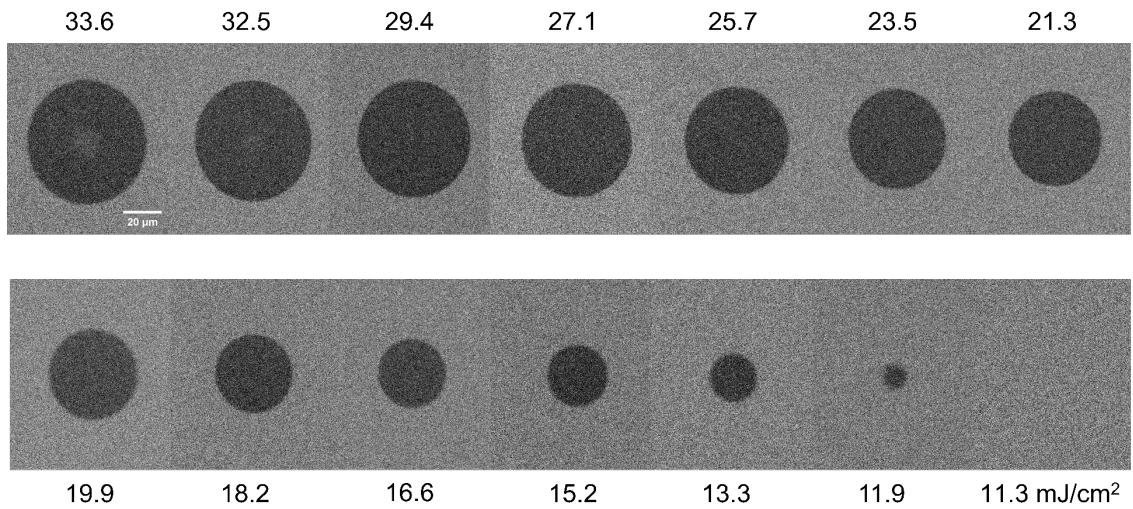

**Fig. S5** | Kerr images of IrMn (5)/Co$_{77}$Gd$_{23}$ (4) after shine to a single linearly polarized laser pulse with various fluences ranging from 11.3 to 33.6 mJ/cm$^2$.

**Section 6. Relative exchange bias changes at various IrMn thicknesses.**

After one single laser pulse, we obtain the maximum exchange bias switched value at each IrMn thickness. And we make a comparison with their initial exchange bias and calculate the relative exchange bias switching rate by $\left|H_e^{\text{switch}}/H_e^{\text{initial}}\right|$. We can note that the thicker the IrMn layer, the less switched exchange bias.

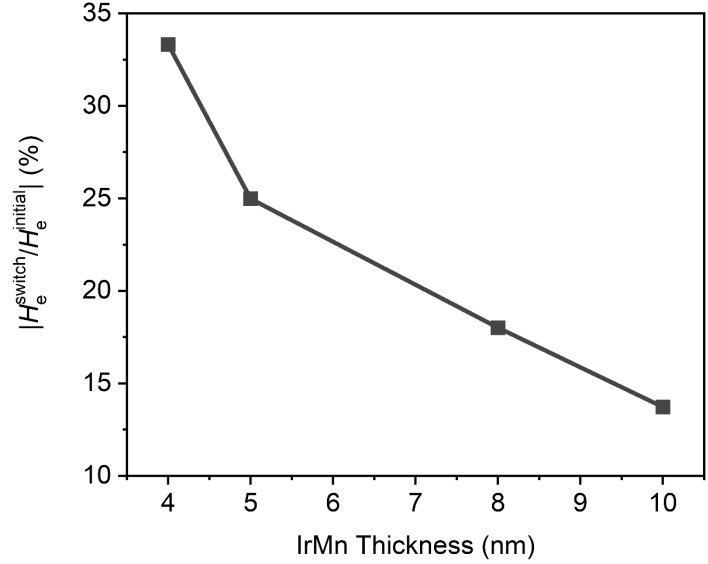

**Fig. S6** | Thickness dependence of the relative exchange bias switching rate $\left|H_e^{\text{switch}}/H_e^{\text{initial}}\right|$.

**Section 7. Evolution of the coercivity as a function of the pulse numbers.**

Fig. S7 shows the evolution of the coercivity $\mu_0 H_c$ of IrMn (5)/Co$_{77}$Gd$_{23}$ (4) as a function of the pulse numbers. As the number of pulses increases, the coercivity tends to decrease and stabilize.

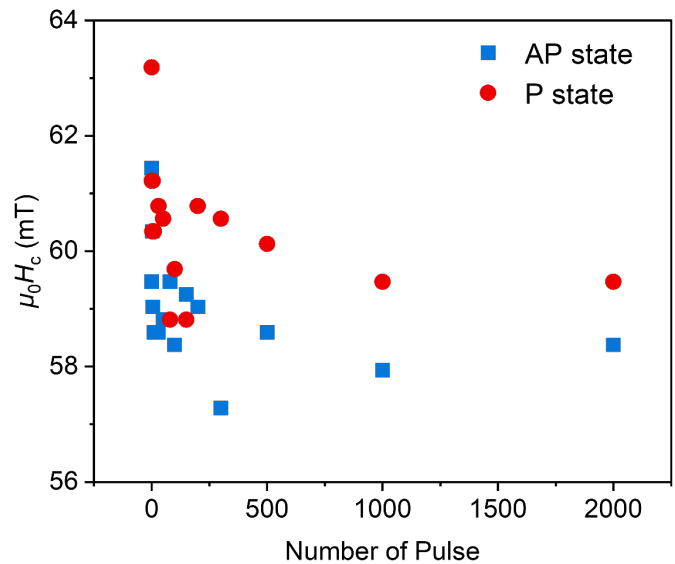

**Fig. S7** | Evolution of the coercivity $\mu_0 H_c$ as a function of the laser pulse numbers.

**Section 8. Exchange bias manipulation by low-intensity multiple femtosecond lasers.**

In IrMn (8)/Co$_{77}$Gd$_{23}$ (4) samples, we apply the single laser pulse with a fluence of 21.5

mJ/cm². Then we use 12.4 mJ/cm² low fluence laser to shine the switching region when the exchange bias field and CoGd magnetization are in parallel (P) and antiparallel (AP) states. The exchange bias field $\mu_0 H_e$ recovers and stabilizes as the number of pulses increases, as shown in Fig. S8a, but still cannot reach the initial state. The coercivity $\mu_0 H_c$ was reduced with the increased pulse number, as shown in Fig. S8b.

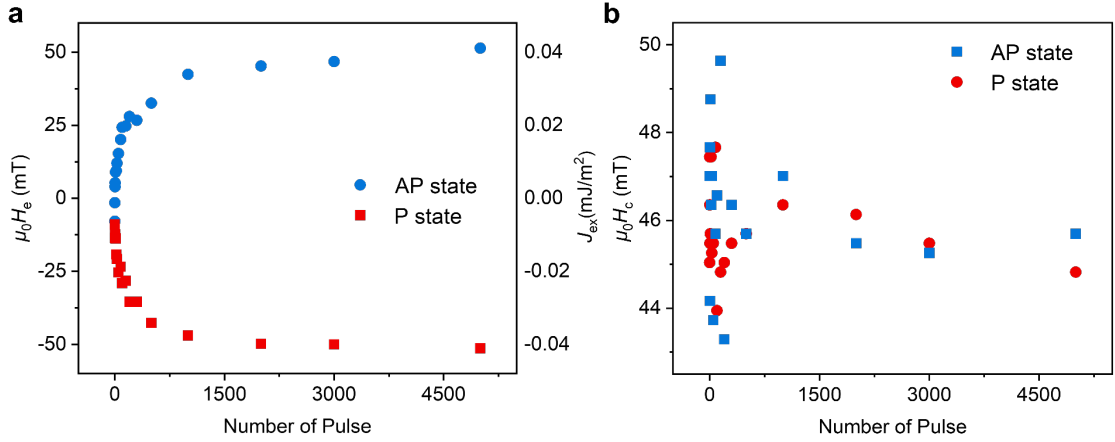

**Fig. S8 | a-b,** After the single-short switching, the switching region is shined by multiple pulses with 12.4 mJ/cm². The exchange bias field $\mu_0 H_e$ recovered along with the CoGd direction, and the coercivity $\mu_0 H_c$ was reduced with the increased pulse number.

**Section 9: CoGd switching dynamics at various IrMn thicknesses.**

The demagnetization and switching dynamics after a 40 fs laser pulse shinned on the annealed IrMn($t_{IrMn}$)/Co$_{77}$Gd$_{23}$(4) ($t_{IrMn}$ =2, 8 nm) samples have been studied as a function of time for different laser fluence (Fig. S9). A demagnetization time of ~ 700 fs could be measured, independent of the applied field. The full reversal requires a typical time of ~ 5 ps.

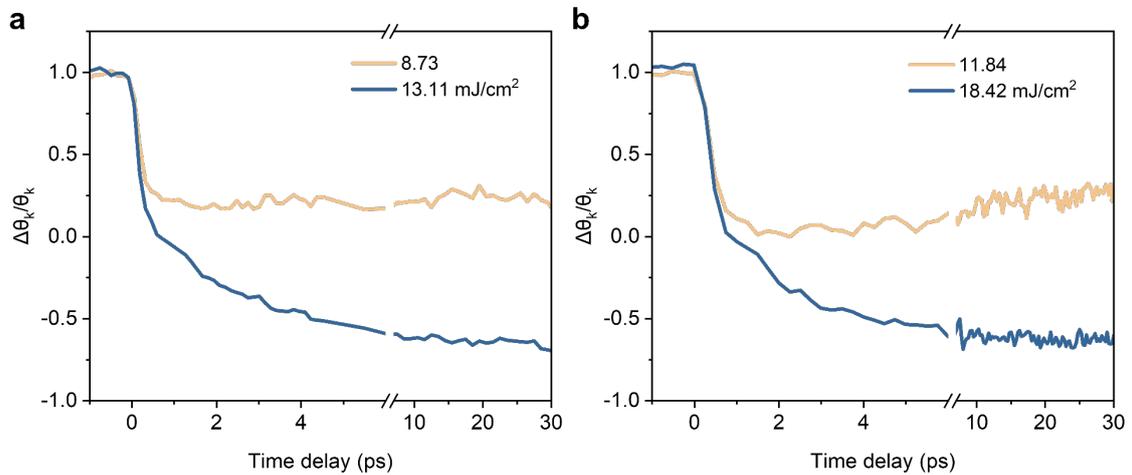

**Fig. S9 | a-b,** CoGd switching dynamics in IrMn(x)/Co$_{77}$Gd$_{23}$(4) (x=2, 8) under an applied field of -220 mT. The demagnetization time is ~ 700 fs and the magnetization switching time is ~ 5 ps.

**Section 10: Parameters of the atomistic simulation.**

| Quantity | Material | Value | Units |
|---|---|---|---|
| NN exchange | Mn | $-6.4 \times 10^{-21}$ | J |
| NNN exchange | Mn | $5.1 \times 10^{-21}$ | J |
| Interlayer exchange | Mn-Co | $1.2 \times 10^{-21}$ | J |
| Interlayer exchange | Mn-Gd | $2.4 \times 10^{-22}$ | J |
| NN exchange | Co-Co | $3.9 \times 10^{-21}$ | J |
| NN exchange | Gd-Co | $-1.25 \times 10^{-21}$ | J |
| NN exchange | Gd-Gd | $1.26 \times 10^{-21}$ | J |
| Neel pair anisotropy | Mn | $-4.22 \times 10^{-22}$ | J |
| Anisotropy Constant | Co | $1.865 \times 10^{-23}$ | J |
| Anisotropy Constant | Gd | $4.035 \times 10^{-24}$ | J |
| Magnetic moment | Mn | 2.6 | μB |
| Magnetic moment | Co | 1.61 | μB |
| Magnetic moment | Gd | 7.63 | μB |
| Gilbert damping | Mn | 0.08 | … |
| Gilbert damping | Co | 0.05 | … |
| Gilbert damping | Gd | 0.02 | … |

**Supplementary Table 1:** Parameters for the atomistic simulation.

**Section 11: Details of the atomistic simulation and field cooling procedure**

To initialise the system in an exchange biased state, we perform a field-cooling procedure, starting above the blocking temperature of the IrMn T = 650 K and cooling to above the compensation point, just below room temperature at T = 240 K. A large setting field of B = 10 T is applied to align the Co sublattice and interfacial Mn moments in the +z direction. The simulation is performed by a Monte Carlo simulation with adaptive moves and a total of $10^8$ steps to allow the system to achieve an equilibrium state at all temperatures. The temperature-dependent magnetization of each material in the system is provided by the field cooling atomistic simulation with an external field as -10 T. From the results in Figure. S10, the Curie temperature of the CoGd is much larger than the IrMn due to the small grain size and enhanced finite size effects[5]. After the field cooling procedure, the system is relaxed in zero field before simulated illumination with a laser pulse.

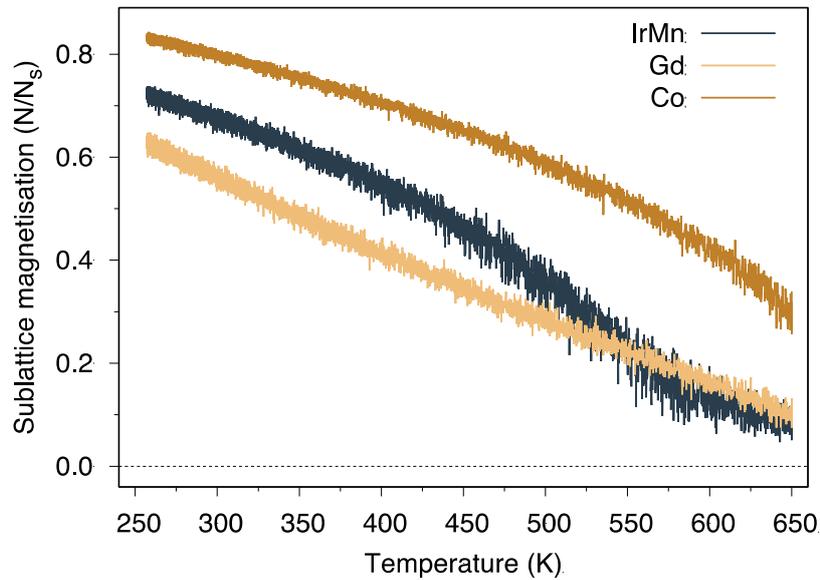

**Fig. S10** | Field cooling Monte Carlo (MC) simulation showing magnetic ordering of the system and Néel temperature of 600 K during the field-cooling procedure over $8\times10^6$ steps.

To show the setting of the grain we show the temperature dependent sublattice magnetization for grain 1 during the field-cooling procedure in Fig. S11. Here there are 8 possible ground states of the magnetization corresponding to the different easy axes of the cubic crystal[6]. In the (111) crystal orientation one of the sublattices aligns along the $\pm z$ direction, corresponding to the degree of coupling with the Gd or Co sublattice. For grain 1, this corresponds to the sublattice with the largest number of moments at the IrMn/CoGd interface aligning anti-parallel with the Co, indicating stronger coupling to the Gd sublattice. In this case, it is sublattice 3, bottom left in Fig. S11.

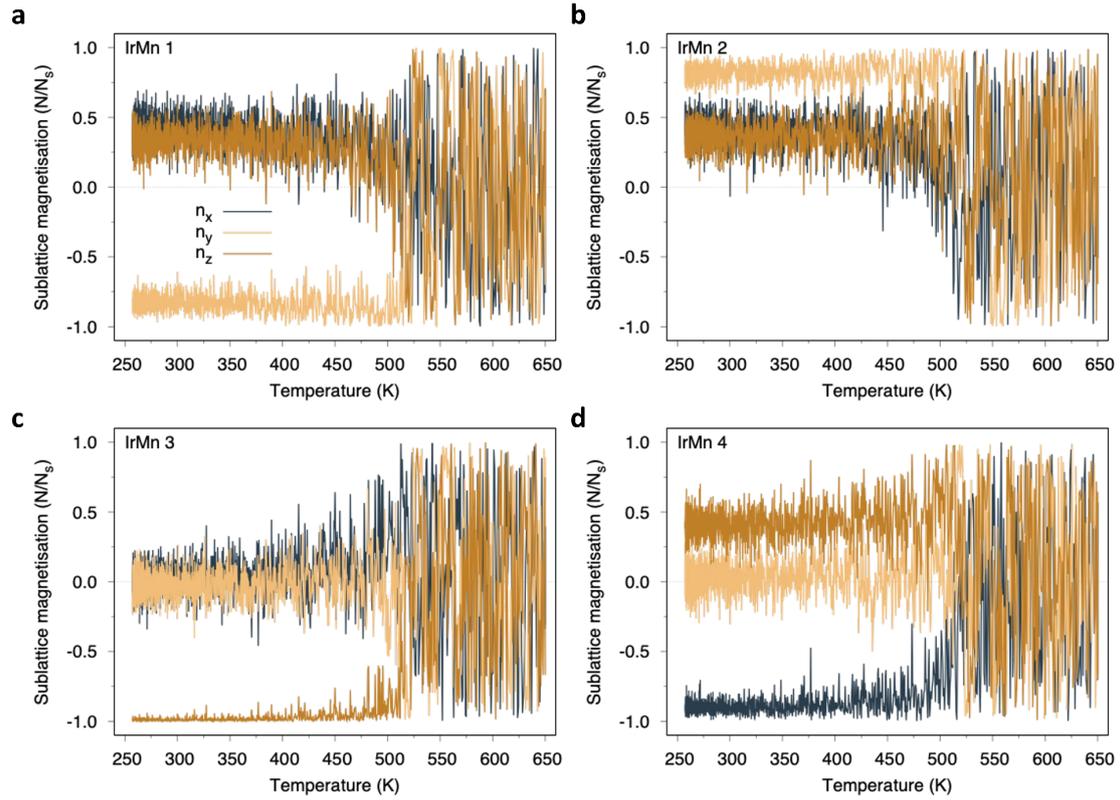

**Fig. S11 | a-d,** Field cooling simulations of a single 5 nm IrMn grain for each of the four magnetic sublattices, showing a stable configuration for T < 500K, suggesting a blocking temperature around T = 500K for nanosecond timescales.

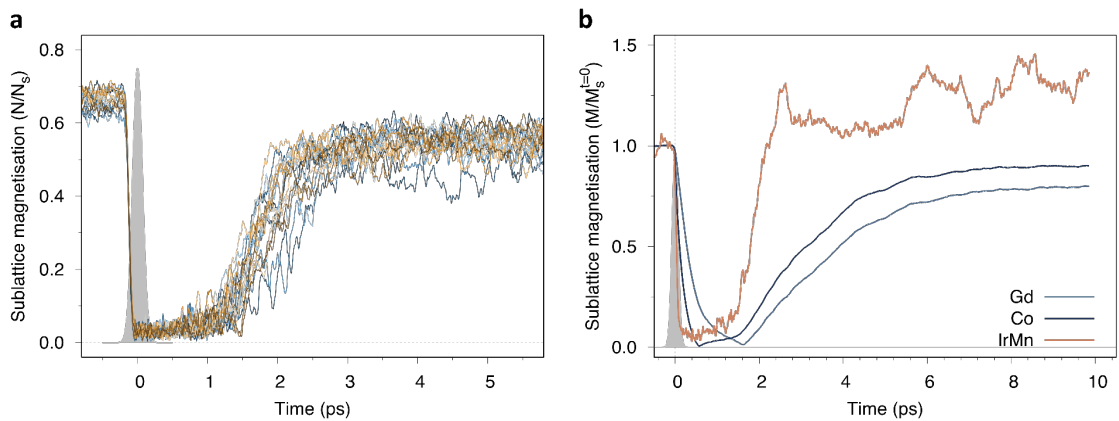

**Fig. S12 | a,** Ultrafast demagnetization and recovery of the IrMn magnetization on a grain wise basis. **b,** Comparative recovery of the Co and Gd magnetization and net magnetic moment in the IrMn (polarised by the CoGd layer). The grain magnetization recovers very quickly after the laser pulse due to the intrinsically fast dynamics of the IrMn, while the recovery of the Co and Gd is slower as it undergoes all-optical switching.

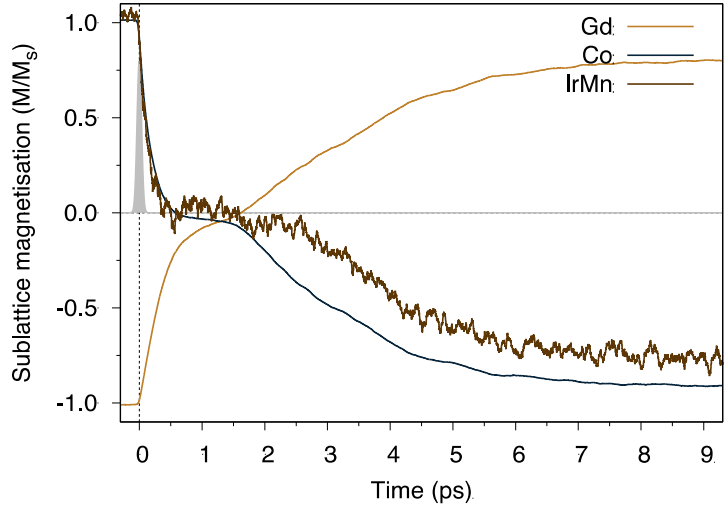

**Fig. S13** | Ultrafast switching of the net magnetization of the Gd, Co and IrMn sublattices. The CoGd reverses by thermally induced switching caused by laser heating, while the reversible Mn moment (polarised by the Co) follows the Co magnetisation on a slightly longer timescale than the switching.

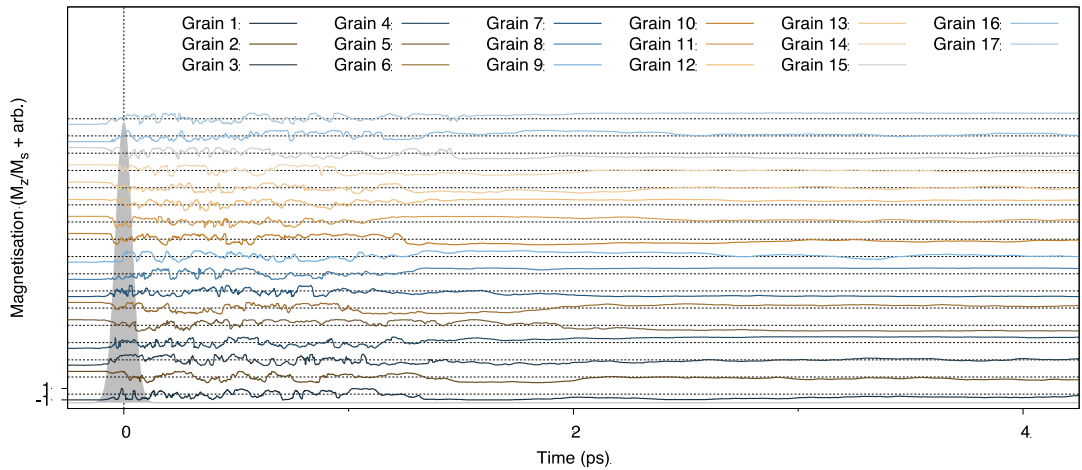

**Fig. S14** | Ultrafast switching of the perpendicular moment of each grain in the system.

Here the extended data for all grains is shown, while the main Fig. 4 shows only the first 5 grains. The data indicate a coherent ultrafast switching process on the short timescale ~ 1.5 ps, but after this initial switching many grains form a multidomain state indicating a switched interface but re-nucleation of the grain order in a different magnetic state. This reduces the effective switching probability due to the formation of an indeterminate orientation of the grains, with $n_z \sim 0$ after the pulse. The data shows all grains form a single domain state but not in a well-defined energy minimum, thus the final state will be random and these grains will not contribute to the exchange bias after the pulse. As the IrMn responds extremely quickly to the laser reaching a spin temperature above the Néel temperature, there are no progressive effects with sequential laser pulses. Thus, the exchange bias is set by the Co direction of each pulse at t = 1.5 ps after the pulse.

| Grain number | Switch after pulse 1 (1.5 ps) | Switch after pulse 1 (long timescale) | Switch after pulse 2 (1.5 ps) | Switch after pulse 2 (long timescale) |
|---|---|---|---|---|
| 1 | N | N | N | Indeterminate |
| 2 | Y | Y | N | N |
| 3 | N | Indeterminate | Y | Indeterminate |
| 4 | Y | Y | Y | N |
| 5 | N | Y | Y | N |
| 6 | Y | Indeterminate | N | N |
| 7 | N | N | Y | Y |
| 8 | Y | Y | Y | N |
| 9 | Y | Indeterminate | N | N |
| 10 | Y | Indeterminate | Y | Y |
| 11 | N | Indeterminate | N | Indeterminate |
| 12 | Y | Indeterminate | Y | Indeterminate |
| 13 | Y | Indeterminate | N | Y |
| 14 | Y | Y | N | Indeterminate |
| 15 | Y | Y | Y | Y |
| 16 | Y | Indeterminate | N | Y |
| 17 | Y | Y | Y | Y |
| Switching probability | 70.5 % | 41.2 % | 53 % | 35.3 % |

**Supplementary Table 2:** Switched and final state of the grains after the first and second laser pulses. The final switching probability represents the exchange bias change at different times after the pulse, taking into account the indeterminate grains. There is no particular correlation between switched and unswitched grains suggesting a random process for each grain switching. The small sample

size of 17 grains gives a large error estimate $\sim \sqrt{17} \sim 25\%$ on the switching probability and so we believe that the process is independent of pulse number, i.e. all grains lose the previous configuration information after each pulse but a certain percentage of grains align with the Co/Gd magnetization after each pulse.